\documentclass[11pt,a4paper]{article}
\usepackage{amsmath,amssymb,graphicx}
\usepackage{hyperref}
\usepackage{geometry}
\usepackage[greek, english]{babel}
\usepackage{alphabeta}
\usepackage{textgreek}
\usepackage{tikz}
\usepackage{cite}
\title{\textbf{Cosmological Tensions as Consistency Conditions for $f(Q)$ Gravity}}
\author{Amare Abebe\\
\small Centre for Space Research, North-West University, South Africa \and
National Institute for Theoretical and Computational Sciences (NITheCS), South Africa\and
Amare.Abebe@nithecs.ac.za}
\date{}

\begin{document}

\maketitle

\begin{abstract}
Cosmology has entered a precision era in which discrepancies between independent datasets, most notably the $H_0$ and $S_8$ tensions, have become robust and statistically significant. These tensions are no longer isolated anomalies but increasingly appear as global consistency constraints on the underlying cosmological model, defining what we will refer to here as a \emph{consistency triangle} of background expansion ($H_0$), structure-growth amplitude ($S_8$), and the redshift-dependence of growth - summarised by the growth index $\gamma$, or equivalently the shape of $f\sigma_8(z)$. The third vertex is non-trivial because in modified-gravity scenarios with a redshift-dependent effective gravitational coupling, growth amplitude and growth shape evolve independently, breaking the rigid coupling characteristic of $\Lambda$CDM. In this work, we use $f(Q)$ gravity as a test case for this emerging paradigm. By drawing on a focused set of recent Bayesian and dynamical-system analyses of the three best-studied functional families - power-law, exponential, and logarithmic - we show that while $f(Q)$ models can alleviate individual tensions, the requirement of simultaneous consistency across $H_0$, $S_8$ and the growth index severely restricts the viable parameter space. A bulk-viscous extension is then briefly examined as a representative illustration of how additional matter-sector freedom is constrained by the same consistency requirement. Our reading of the current literature supports the view that cosmological tensions should be interpreted as global consistency conditions, and that viable extensions of $\Lambda$CDM must satisfy this multi-probe constraint \cite{CosmoVerse,DiValentino2025Corfu}. Within this framework, only a restricted subset of $f(Q)$ models remains competitive.
\end{abstract}

\section{Cosmology in the Precision-Tension Era}

The standard $\Lambda$CDM model has achieved remarkable success in describing the large-scale universe, particularly in its agreement with CMB observations \cite{Planck2018}. However, a growing body of observational evidence now challenges its completeness.

The Hubble constant tension is the sharpest such challenge. The early-Universe inference from \emph{Planck} 2018, $H_0 = 67.27 \pm 0.60~\mathrm{km\,s^{-1}\,Mpc^{-1}}$ \cite{Planck2018}, sits in $5$ - $6\sigma$ disagreement with late-universe distance-ladder values around $H_0 \approx 73~\mathrm{km\,s^{-1}\,Mpc^{-1}}$, with strong-lensing time delays providing an additional, partly independent late-universe anchor \cite{Birrer2025}. The $S_8$ tension reveals a related mismatch in the amplitude of structure growth between CMB and weak-lensing surveys, and it is increasingly framed in terms of a suppressed late-time growth index $\gamma$ inconsistent with the $\Lambda$CDM expectation $\gamma \simeq 0.55$.

As emphasised in the Corfu meeting report~\cite{DiValentino2025Corfu}, cosmology has now firmly entered a \textit{precision-tension era}, in which these discrepancies are robust features of the data; the hypothesis that they arise purely from unknown systematics is no longer dominant, as also stressed in the CosmoVerse White Paper~\cite{CosmoVerse}.

Recent large-scale structure observations further sharpen this picture. DESI DR2 provides high-precision BAO measurements broadly consistent with $\Lambda$CDM while leaving room for late-time extensions \cite{DESI2025}, whereas joint analyses of ACT, SPT, and \emph{Planck} CMB-lensing data show strong consistency in structure-growth measurements \cite{Qu2026}. Together, these results impose stringent constraints on any proposed resolution of cosmological tensions.

A key conceptual shift has therefore emerged: viable cosmological models must satisfy \textit{global consistency} across multiple probes, including the expansion history, the present-day amplitude of structure growth, and its redshift-dependence \cite{CosmoVerse,DiValentino2025Corfu}. We formalise this requirement below as the \emph{consistency triangle} - expansion ($H_0$), growth amplitude ($S_8$ or $\sigma_8$), growth shape (encoded in the growth index $\gamma$ or in the redshift-evolution of $f\sigma_8(z)$) - and use it as the organising principle for the rest of this paper. The three vertices are not redundant: in modified-gravity scenarios with a redshift-dependent effective gravitational coupling, the present-day amplitude of growth and its evolution with redshift respond differently to changes in the underlying theory, so requiring simultaneous consistency at all three vertices is a strictly stronger constraint than requiring consistency at any pair of them.

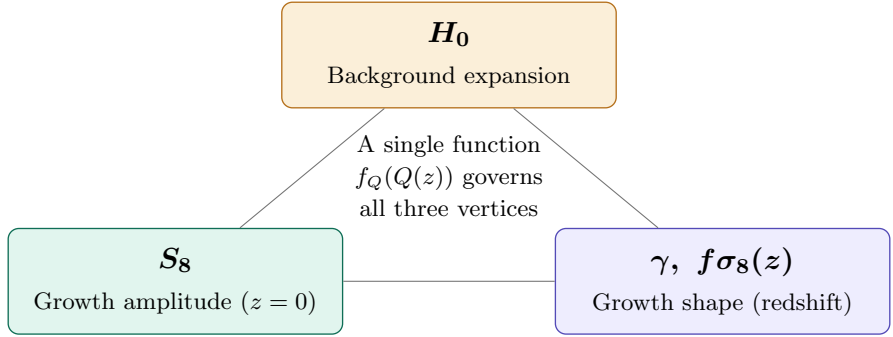
\begin{figure}[t]
\centering
\definecolor{vAmberFill}{HTML}{FBEFD9}
\definecolor{vAmberLine}{HTML}{B5701B}
\definecolor{vTealFill}{HTML}{E1F5EE}
\definecolor{vTealLine}{HTML}{0F6E56}
\definecolor{vPurpFill}{HTML}{EEEDFE}
\definecolor{vPurpLine}{HTML}{534AB7}
\begin{tikzpicture}[
  vertex/.style={
    rounded corners=4pt, draw, line width=0.5pt,
    minimum width=4.4cm, minimum height=1.4cm,
    align=center, inner sep=5pt
  },
  triedge/.style={black!55, line width=0.4pt}
]
\node[vertex, fill=vAmberFill, draw=vAmberLine] (H0) at (0, 3.0)
  {{\bfseries\boldmath$H_0$}\\[2pt]{\footnotesize Background expansion}};
\node[vertex, fill=vTealFill, draw=vTealLine] (S8) at (-3.6, 0)
  {{\bfseries\boldmath$S_8$}\\[2pt]{\footnotesize Growth amplitude ($z=0$)}};
\node[vertex, fill=vPurpFill, draw=vPurpLine] (gam) at (3.6, 0)
  {{\bfseries\boldmath$\gamma,\ f\sigma_8(z)$}\\[2pt]{\footnotesize Growth shape (redshift)}};
\draw[triedge] (H0) -- (S8);
\draw[triedge] (H0) -- (gam);
\draw[triedge] (S8) -- (gam);
\node[align=center, font=\footnotesize] at (0, 1.4)
  {A single function\\[1pt]$f_Q(Q(z))$ governs\\[1pt]all three vertices};
\end{tikzpicture}
\caption{The cosmological consistency triangle. Any viable cosmological model must simultaneously match three constraints: the background expansion ($H_0$), the present-day amplitude of structure growth ($S_8$ or $\sigma_8$), and the redshift-evolution of growth - encoded in the growth index $\gamma$ or equivalently in the shape of $f\sigma_8(z)$. The three vertices are not independent observational \emph{statements}, since they are different functionals of the same underlying matter-density field; but in modified-gravity scenarios with a redshift-dependent effective gravitational coupling they become independent \emph{constraints on theory}. In $f(Q)$ gravity, all three are governed by the single function $f_Q(Q(z))$: $f_Q(Q_0)$ controls the present-day growth amplitude through $G_{\rm eff}(z=0) = G_{\rm N}/f_Q(Q_0)$, while the full redshift profile $f_Q(Q(z))$ controls the growth shape.}
\label{fig:triangle}
\end{figure}
\section{$f(Q)$ Gravity as a Test Case of Consistency}

\subsection{Why $f(Q)$ gravity}

Modified gravity provides a natural framework for exploring deviations from $\Lambda$CDM. Among the recently active geometric alternatives, $f(Q)$ gravity  -  the non-linear extension of the symmetric teleparallel equivalent of General Relativity (STEGR) - has emerged as one of the most versatile frameworks for cosmology. In this class of theories, gravitation is encoded in the non-metricity scalar $Q$ rather than curvature $R$ or torsion $T$, while the field equations remain second order, which makes the framework technically attractive for model building and data confrontation. Recent reviews emphasise that $f(Q)$ gravity has rapidly developed into a broad research programme spanning late-time acceleration, early-universe dynamics, perturbations, compact objects, and cosmological tensions \cite{Heisenberg2024Review}.

From the point of view of cosmological tensions, the appeal of $f(Q)$ gravity is twofold: it modifies both the background expansion history, which is relevant to the $H_0$ discrepancy, and the effective gravitational coupling governing structure formation, which is relevant to the $S_8$ and more general growth tensions. Recent dedicated analyses show that some $f(Q)$ models can shift the inferred $H_0$ upwards relative to \emph{Planck} $\Lambda$CDM, while selected sectors with effectively weakened gravity, $G_{\rm eff}<G_{\rm N}$, can suppress structure growth enough to move predictions towards weak-lensing-preferred values \cite{Boiza2025}. At the same time, DESI-era BAO measurements and CMB-lensing-derived growth measurements make it clear that no proposed modification can be assessed in isolation: viability requires simultaneous consistency of background expansion, growth amplitude, and growth shape \cite{CosmoVerse,DiValentino2025Corfu,DESI2025,Qu2026}.

\subsection{Field equations and the $H_0$ - $S_8$ link}

Throughout this paper we adopt the spatially flat FLRW geometry and the \emph{coincident gauge}, in which the affine-connection coefficients vanish in Cartesian coordinates and the non-metricity scalar reduces to $Q = 6 H^2$, with $H \equiv \dot a/a$. This choice corresponds to ``Connection~I'' in recent classifications of admissible homogeneous and isotropic connections in $f(Q)$ cosmology; alternative non-coincident branches yield modified dynamics for the same functional form $f(Q)$ \cite{Ayuso2025Connection,Pradhan2024NonCoincident} and lie outside the scope of this proceedings. Variation of the action 
\begin{equation}
S=\int d^4x\sqrt{-g}\,[f(Q)/(2\kappa^2) + \mathcal{L}_m]\;,
\end{equation}
 with $\kappa^2 = 8\pi G_{\rm N}$ gives the modified Friedmann equation
\begin{equation}
6\,f_Q\, H^2 \;-\; \tfrac{1}{2}\,f(Q) \;=\; \kappa^2\, \rho_{\rm m}\,,
\label{eq:Friedmann}
\end{equation}
where $f_Q \equiv df/dQ$. STEGR, and hence $\Lambda$CDM with a bare cosmological constant, is recovered in the limit $f(Q) \to Q - 2\Lambda$.

At linearly perturbed level, in the quasi-static, sub-horizon limit relevant for $f\sigma_8$ and weak-lensing data, the matter density contrast obeys $\ddot\delta_{\rm m} + 2H\,\dot\delta_{\rm m} - 4\pi\, G_{\rm eff}(z)\,\rho_{\rm m}\,\delta_{\rm m} = 0$, with the model-dependent effective gravitational coupling \cite{Sahlu2025,Khyllep2022DynSys,KhyllepGrowth2021}
\begin{equation}
\frac{G_{\rm eff}(z)}{G_{\rm N}} \;=\; \frac{1}{f_Q\!\left(Q(z)\right)}\,.
\label{eq:Geff}
\end{equation}
Equation~\eqref{eq:Geff} encodes the central mechanism by which $f(Q)$ gravity acts on the structure-growth sector: branches with $f_Q > 1$ at late times realise $G_{\rm eff} < G_{\rm N}$ and suppress structure growth, while branches with $f_Q < 1$ enhance it. The same function $f_Q$, evaluated at the background level, controls the modification of the expansion history through equation~\eqref{eq:Friedmann}. Crucially for what follows, $f_Q$ is \emph{redshift-dependent}: the present-day value $f_Q(Q_0)$ effectively sets the growth amplitude (and hence $S_8$), while the full redshift profile $f_Q(Q(z))$ sets the growth shape (and hence the growth index $\gamma$ or, equivalently, $f\sigma_8(z)$). Amplitude and shape are therefore controlled by \emph{different aspects of the same modification}, rather than being slaved together as in $\Lambda$CDM. This is precisely what makes $f(Q)$ gravity a sharp, three-axis test case of the consistency triangle: a single function $f_Q$ governs all three vertices, but the three vertices are nonetheless independently constrained by data.

\subsection{Three representative functional families}

The literature on $f(Q)$ cosmology is now sufficiently broad that it is useful to organise it by classes of functional forms. Throughout this paper, the three functionals we focus on are
\begin{align}
f_1(Q) &= Q + \alpha\,Q^n, \label{eq:f1}\\
f_2(Q) &= Q + \beta\,Q_0\!\left(1 - e^{-p\sqrt{Q/Q_0}}\right), \label{eq:f2}\\
f_3(Q) &= Q + \epsilon\,\ln\!\left(\Gamma\, Q/Q_0\right), \label{eq:f3}
\end{align}
where $\alpha,\beta$ are dimensionless deformation amplitudes, $\epsilon$  is a parameter with the dimensions of $Q$ that can be derived from the cosmological observables, $n$ and $p$ are exponents controlling the steepness of the deviation from STEGR, $Q_0$ denotes the present-day value of the non-metricity scalar, and $\Gamma$ is a normalisation parameter chosen so that the logarithmic correction vanishes at a chosen reference epoch. They span three of the most important and best-studied sectors of the theory space: minimal algebraic deformations, smooth exponential modifications, and mild logarithmic corrections, and they recur across most recent observational analyses, allowing direct cross-comparison.

\paragraph{Power-law deformations.}
The simplest class, equation~\eqref{eq:f1}, provides a minimal one-function departure from STEGR. It has been analysed at both background and perturbation level, including with OHD, Pantheon$+$, RSD and $f\sigma_8$ data, and it remains among the most competitive sectors of the theory space \cite{Sahlu2025,Khyllep2022DynSys,KhyllepGrowth2021,Mhamdi2024PowerExp,Shabani2024NonFlat}. In the joint expansion-plus-growth analysis of \cite{Sahlu2025}, the power-law model is precisely the class that remains observationally viable once both kinds of probes are combined.

\paragraph{Exponential deformations.}
The exponential class, equation~\eqref{eq:f2}, is usually designed to recover STEGR at high redshift while introducing a smooth late-time deformation capable of mimicking dark energy. Recent observational studies have directly compared power-law and exponential sectors and, interestingly, found cases in which the exponential branch is statistically favoured over both the power-law case and $\Lambda$CDM for specific dataset combinations \cite{Mhamdi2024PowerExp}. Closely related work has proposed novel exponential and hyperbolic-tangent-inspired $f(Q)$ constructions tailored to late-time data and standard-siren forecasts, with one such model reported to reduce the Hubble tension substantially when electromagnetic data are combined with simulated Einstein Telescope observations \cite{Su2025ETSirens}.

\paragraph{Logarithmic deformations.}
Logarithmic corrections, equation~\eqref{eq:f3}, or equivalently $f(Q)=Q/(8\pi G_{\rm N})-\alpha\ln(Q/Q_0)$, are typically motivated as mild, slowly varying deformations of STEGR that mimic late-time acceleration geometrically. Recent work has constrained logarithmic models with local SNIa and BAO data, energy conditions, mock standard sirens, and growth observables \cite{Najera2023LogGW,Gadbail2024ThreeModels,KhyllepGrowth2021}: they remain phenomenologically interesting even if they are typically less competitive than the best power-law scenarios, and they predict measurable deviations from GR in the gravitational-wave luminosity-distance ratio that future detectors will be able to test.

\subsection{Wider landscape: connection-sensitive and extended sectors}

The recent literature has also shown that the cosmology of $f(Q)$ gravity is structurally richer than early coincident-gauge studies suggested. In non-flat FLRW geometry, keeping $f(Q)$ arbitrary already yields new curvature-driven critical points, including accelerating phases and states that can alleviate the coincidence problem \cite{Shabani2024NonFlat}. Non-coincident formulations have been used to reanalyse familiar power-law, exponential and logarithmic corrections under a non-trivial affine connection, leading to modified Friedmann dynamics and fresh observational constraints \cite{Pradhan2024NonCoincident}. 

Non-coincident formulations have been used to reanalyse familiar power-law, exponential and logarithmic corrections under a non-trivial affine connection, leading to modified Friedmann dynamics and fresh observational constraints \cite{Pradhan2024NonCoincident,Paliathanasis2025DESI}; in particular, the recent DESI DR2$\,+\,$GRBs analysis of \cite{Paliathanasis2025DESI} reformulates the non-coincident power-law model $f(Q) \simeq Q^{n/(n-1)}$ as an effective two-scalar-field, quintom-like dark-energy theory and finds a best-fit $n \simeq 0.33$, with a smaller $\chi^2_{\min}$ than $\Lambda$CDM and weak Bayesian evidence in favour of $f(Q)$ once gamma-ray bursts are included. The minisuperspace analysis of De and Paliathanasis~\cite{DePaliathanasis2025} complements this picture by deriving exact non-coincident $f(Q)$ solutions - including de Sitter, scaling, $\Lambda$CDM-mimicking, Chaplygin-gas and CPL branches - in closed analytic form, showing that the non-coincident sector admits a substantially richer solution space than the coincident gauge. 
The physical relevance of the affine connection itself has been highlighted further: different connection choices compatible with homogeneous and isotropic cosmology can generate genuinely different cosmological evolutions for the same functional form, including Born-Infeld-type cases that smooth Big-Bang or Big-Rip singularities into de Sitter phases \cite{Ayuso2025Connection}. Beyond pure $f(Q)$, several extensions enlarge the available phenomenology - $f(Q,\mathcal{L}_m)$ \cite{Hazarika2025fQLm}, $f(Q,T)$, $f(Q,\mathcal{C})$, viscous or effective-fluid versions \cite{SahluViscous2025} -  as well as emergent-universe and singularity-resolution scenarios \cite{Shabani2025Emergent}. The common theme is that the flexibility of the non-metricity sector allows one to engineer deformations that are mild at early times yet non-trivial at late times, which is precisely the regime where the current tension network is most constraining.

\section{What Recent Analyses Tell Us}
\label{sec:landscape}

Rather than reproducing posteriors here, we anchor the discussion of the consistency triangle in the quantitative results that have already appeared for the three functional families above. The picture that emerges is consistent across independent analyses but more nuanced than any single-dataset confrontation suggests.

\paragraph{Background expansion and $H_0$.}
The recent Bayesian study of three representative $f(Q)$ models by Boiza, Petronikolou, Bouhmadi-L\'opez and Saridakis~\cite{Boiza2025}, using cosmic chronometers, Type-Ia supernovae, gamma-ray bursts, BAO and CMB distance priors, finds that most of the analysed models yield $H_0$ values higher than the \emph{Planck} $\Lambda$CDM baseline of $H_0 = 67.27 \pm 0.60~\mathrm{km\,s^{-1}\,Mpc^{-1}}$ \cite{Planck2018}, achieving a partial alleviation of the $H_0$ tension. The same paper identifies one branch satisfying $G_{\rm eff} < G_{\rm N}$ that simultaneously predicts $S_8$ values compatible with weak-lensing measurements, but at the price of mild internal inconsistencies between subsets of data, which limit the overall statistical preference relative to $\Lambda$CDM. This trade-off is, in our view, the most directly visible manifestation of the consistency triangle in the current $f(Q)$ literature.

\paragraph{Power-law versus exponential at the perturbation level.}
The Bayesian analysis of Mhamdi et al.~\cite{Mhamdi2024PowerExp}, using Pantheon$^+$, $H(z)$ and RSD data and both AIC and BIC criteria, reports that the exponential $f(Q)$ branch is statistically favoured over both the power-law case and $\Lambda$CDM for that specific dataset combination. Conversely, when the dataset is enlarged to include CMB distance priors and gamma-ray bursts as in~\cite{Boiza2025}, no $f(Q)$ branch decisively beats $\Lambda$CDM. The contrast between these conclusions is itself instructive: a single AIC/BIC verdict on a partial dataset does not survive the move to a more complete probe combination - which is exactly the message of the consistency-triangle viewpoint.

\paragraph{Growth amplitude and growth shape.}
The growth-of-structure analysis of Sahlu, de la Cruz-Dombriz and Abebe~\cite{Sahlu2025}, performed in the $1{+}3$ covariant formalism with $f\sigma_8$, $f$ and $\sigma_8$ measurements from RSD, VIPERS and SDSS, confirms that the power-law family $f(Q)=Q+\alpha Q^n$ remains observationally viable once expansion and growth data are jointly imposed - but the allowed exponent $n$ is forced very close to zero, that is, very close to STEGR. The growth-shape vertex is constrained more directly by the dedicated growth-index analysis of Mhamdi et al.~\cite{Mhamdi2024GrowthIndex}, who report $\gamma = 0.571^{+0.095}_{-0.110}$ within $f(Q)$ gravity, slightly above the $\Lambda$CDM value $\gamma \simeq 0.55$ but well below the suppressed-growth values $\gamma \approx 0.63$ - $0.64$ at which late-time growth-suppression analyses claim to alleviate the $S_8$ tension. This places $f(Q)$ gravity in an awkward middle position along the growth-shape vertex: its preferred $\gamma$ is consistent with $\Lambda$CDM and with a partial $S_8$ alleviation, but cannot reach the more strongly suppressed values needed to fully resolve the growth tension. The dynamical-system analysis of Khyllep et al.~\cite{Khyllep2022DynSys} confirms that both the power-law and the exponential branches admit a matter-dominated saddle point with the correct linear growth rate, followed by a stable de Sitter attractor; on those grounds, neither family is theoretically excluded.

\paragraph{Gravitational-wave and standard-siren forecasts.}
For the logarithmic class, N\'ajera, Ar\'aoz Alvarado and Escamilla-Rivera~\cite{Najera2023LogGW} forecast that mock Einstein-Telescope and LISA standard sirens combined with current SNIa and BAO data could pin down $H_0$ to better than $1\%$ and $\Omega_m$ to better than $6\%$, while predicting a model-induced ratio $d_L^{\,\rm gw}(z)/d_L^{\,\rm em}(z)$ that deviates from unity by more than $13\%$ at $z=1$ and continues to grow with redshift. Su, He and Zhang~\cite{Su2025ETSirens} reach a complementary conclusion across power-law, exponential and hyperbolic-tangent variants: their $f(Q)_{\rm PE}$ mixed-model is already disfavoured by current electromagnetic data, while two hyperbolic-tangent constructions will be excluded outright by Einstein-Telescope-class siren samples. Cross-cutting probes such as standard-siren cosmology and time-delay strong lensing therefore do not define an additional independent vertex of the consistency triangle. They are, instead, simultaneous tests of multiple vertices: time-delay lensing measures $H_0$ directly via $D_{\Delta t}$~\cite{Birrer2025} (the expansion vertex), while the gravitational-wave luminosity-distance ratio $d_L^{\,\rm gw}/d_L^{\,\rm em}$ tracks the redshift evolution of the modified gravitational coupling, which is precisely what fixes the growth-shape vertex through equation~\eqref{eq:Geff}. They are, in this sense, some of the most decisive cross-vertex consistency checks expected in the coming decade.

\paragraph{Tension-resolution attempts.}
Finally, the broader $f(Q)$ tension-resolution programme exemplified by~\cite{Boiza2025,Su2025ETSirens} reaches a nuanced verdict: $f(Q)$ models can alleviate the $H_0$ tension partially, and a narrow branch can simultaneously realise $G_{\rm eff} < G_{\rm N}$ and reduce $S_8$ towards weak-lensing-preferred values, but no single model in the literature yet provides a fully simultaneous, statistically preferred resolution along all three vertices ($H_0$, $S_8$, $\gamma$) of the consistency triangle.

\section{An Illustrative Extension: Bulk Viscosity}

To illustrate how additional matter-sector freedom is judged against the same consistency requirement, we briefly examine an Eckart-type bulk-viscous extension in which the effective pressure is $p_{\rm eff} = p - 3\,\zeta(H)\,H$ with the Hubble-rate parameterisation $\zeta(H) = \zeta_0 + \zeta_1\,H/H_0$, following~\cite{SahluViscous2025}. This is the simplest physically motivated choice that reduces to a constant viscosity for $\zeta_1 \to 0$ and to a strictly Hubble-tracking viscosity for $\zeta_0 \to 0$. The dedicated analysis of~\cite{SahluViscous2025}, which combines DESI BAO, cosmic chronometers, Pantheon$^+$\,SH0ES, $f$ and $f\sigma_8$ data, finds that adding bulk viscosity to the three functional families does suppress clustering and pull $\sigma_8$ downwards, as expected on physical grounds, but does \emph{not} translate into improved AIC/BIC scores once the additional $\zeta_0,\zeta_1$ degrees of freedom are penalised. In the language of the consistency triangle, the viscous extension shifts predictions along the growth-amplitude vertex (lower $\sigma_8$) but pays for it at the parsimony level, without bringing the model closer to simultaneous $H_0$-$S_8$-$\gamma$ consistency. This is a useful negative result: it shows that simply adding dissipative matter-sector freedom does not, on its own, resolve the constraint structure imposed by the consistency triangle.

\section{Discussion: Tensions as Consistency Conditions}

A clear pattern emerges from the analyses surveyed above. $f(Q)$ gravity can shift $H_0$ towards late-universe values, and it can suppress structure growth at late times via $G_{\rm eff} < G_{\rm N}$. However, achieving both simultaneously while also producing a growth-shape (encoded in $\gamma$ or $f\sigma_8(z)$) consistent with the data proves difficult, and remains a narrow corner of parameter space rather than a generic feature.

This reflects a deeper point. Cosmological tensions are not independent problems but interconnected constraints. As emphasised in~\cite{CosmoVerse,DiValentino2025Corfu}, progress now comes from joint analyses across multiple probes. In this framework, tensions define a constrained region of theory space; viable models must satisfy a \textit{consistency triangle} involving:
\begin{itemize}
\item Expansion history ($H_0$);
\item Growth amplitude ($S_8$ or $\sigma_8$ at $z = 0$);
\item Growth shape: redshift-evolution of $f\sigma_8(z)$, summarised by the growth index $\gamma$.
\end{itemize}
The three vertices are not independent observational \emph{statements} - they are different functionals of the same underlying matter-density field - but they are independent \emph{constraints on theory}, because in modified-gravity models with redshift-dependent gravitational coupling they respond to different aspects of the modification. $f(Q)$ gravity illustrates both the promise and the limits of this picture: the right architectural ingredients are present - a single function $f_Q$ controlling background, growth amplitude (via $f_Q(Q_0)$) and growth shape (via $f_Q(Q(z))$) -  but recovering all three vertices simultaneously is non-trivial, and current data already disqualify large parts of the originally permissible parameter space.

\section{Conclusions}

We have analysed $f(Q)$ gravity as a test case of the emerging global-consistency paradigm in cosmology, anchoring the discussion in recent published Bayesian and dynamical-system analyses rather than fresh posteriors. Our main conclusions are:
\begin{itemize}
\item $f(Q)$ models can alleviate individual tensions, particularly $H_0$ \cite{Boiza2025,Sahlu2025};
\item Joint background-plus-growth data impose strong combined constraints, narrowing the viable parameter space considerably and forcing the power-law exponent $n$ close to its STEGR limit;
\item A narrow $G_{\rm eff} < G_{\rm N}$ branch can simultaneously raise $H_0$ and reduce $S_8$, but at the price of mild internal data inconsistencies \cite{Boiza2025};
\item The third vertex of the consistency triangle - the growth shape, encoded in $\gamma$ or $f\sigma_8(z)$ - is non-trivially constrained: $f(Q)$ gravity prefers $\gamma \approx 0.57$ \cite{Mhamdi2024GrowthIndex}, intermediate between the $\Lambda$CDM value and the strongly suppressed values claimed by some late-time growth-suppression scenarios;
\item Bulk-viscous matter-sector extensions illustrate how added freedom is penalised at the AIC/BIC level when it does not improve consistency-triangle alignment \cite{SahluViscous2025};
\item No model currently in the literature provides a complete, statistically preferred simultaneous resolution along all three vertices ($H_0$, $S_8$, $\gamma$) of the consistency triangle.
\end{itemize}
The broader implication is that cosmological tensions should be interpreted as consistency conditions on the underlying theory rather than as independent anomalies. Future surveys - DESI DR3 and beyond, Euclid weak lensing, next-generation CMB experiments such as Simons Observatory and CMB-S4, and Einstein-Telescope-class standard sirens - will be decisive in testing whether modified-gravity frameworks such as $f(Q)$ can satisfy these constraints, or whether a more radical revision of the cosmological model is required \cite{DESI2025,Najera2023LogGW,Su2025ETSirens}.

\bibliographystyle{unsrt}
\bibliography{fq_tensions}

\end{document}